\documentclass{article}

\newcommand{\mc}[1]{\ensuremath{\mathcal{#1}}}
\newcommand{\ket}[1]{\ensuremath{\left|  #1 \right\rangle}}
\newcommand{\bra}[1]{\ensuremath{\left\langle #1 \right|}}
\newcommand{\proj}[2]{\ensuremath{\ket{#1} \bra{#2}}}
\newcommand{\matel}[3]{\ensuremath{\bra{#1} #2 \ket{#3}}}
\newcommand{\op}[1]{\ensuremath{\widehat{\textsf{\ensuremath{#1}}}}}
\newcommand{\be}{\begin{equation}}
\newcommand{\ee}{\end{equation}}

\newcommand{\eg}{\mbox{e.\,g.\,\ }}
\newcommand{\egc}{\mbox{e.\,g.\,}}

\usepackage{chicago}
\usepackage{latexsym}

\newcommand{\crossout}{\!\!\!\!\!\! /}
\newtheorem{lemma}{Lemma}

\begin{document}

\title{Quantum Probability from Subjective Likelihood: improving on
Deutsch's proof of the probability rule}
\author{David Wallace}
\date{April 2005}
\maketitle

\begin{abstract}
I present a proof of the quantum probability rule from 
decision-theoretic assumptions, in the context of the Everett
interpretation. The basic ideas behind the proof are those presented in 
Deutsch's recent proof of the probability rule, but the proof is simpler
and proceeds from weaker decision-theoretic assumptions. This makes it
easier to discuss the conceptual ideas involved in the proof, and to
show that they are defensible.
\end{abstract}

\section{Introduction}

\begin{quote}
The mathematical formalism of the quantum theory is capable of yielding
its own interpretation.

\begin{flushright}Bryce DeWitt 
\citeyear{dewitt}
\end{flushright}
\end{quote}

If I were to pick one theme as central to the tangled development of the
Everett interpretation of quantum mechanics, it would probably be:
\emph{the formalism is to be left alone.} What distinguished Everett's original paper
both from the Dirac-von Neumann collapse-of-the-wavefunction orthodoxy
and from contemporary rivals such as the de Broglie-Bohm theory was its
insistence that unitary quantum mechanics need not be supplemented in
any way (whether by hidden variables, by new dynamical processes, or
whatever).

Many commentators on the Everett interpretation --- even some, like
David Deutsch, who are sympathetic to it \cite{deutsch85} --- have at
various points and for various reasons retreated from this claim.\footnote{It 
is perhaps worth noting that Deutsch no
longer sees any need to modify the formalism, and is now happy with
a decoherence-based solution to the preferred-basis problem.} The
``preferred basis problem'', for instance, has induced many to suppose
that quantum mechanics must be supplemented by some explicit rule that
picks out one basis as physically special. Many suggestions were made
for such a rule (\citeN{barrett} discusses several); all seem to
undermine the elegance (and perhaps more crucially, the relativistic
covariance) of Everett's original proposal. Now the rise of decoherence
theory has produced a broad consensus among supporters of Everett (in
physics, if perhaps not yet in philosophy) that the supplementation was
after all not necessary. (For more details of this story, see
Wallace \citeyear{wallaceworlds,wallacestructure} and references therein.)

Similarly, various commentators (notably \citeN{bell}, \citeN{butterfield} and
\citeN{albertloewer}) have suggested that the Everett interpretation has
a problem with persistence of objects (particles, cats, people, \ldots)
over time, and some have been motivated to add explicit structure to
quantum mechanics in order to account for this persistence (in
particular, this is a prime motivation for Albert and Loewer's original
Many-Minds theory). Such moves again undermine the rationale for an
Everett-type interpretation. And again, a response to such criticism
which does not require changes to the formalism was eventually
forthcoming. It was perhaps implicit in Everett's original discussion
of observer memory states, and in Gell-Mann and Hartle's later notion of
IGUSs \cite{gellmannhartle}; it has been made explicit by Simon
Saunders' work \cite{saundersprobability} on the analogy between Everettian
branching and Parfittian fission.

In both these cases, my point is not that the critics were foolish or
mistaken: Everettians were indeed obliged to come up with solutions both
to the preferred-basis problem and to the problem of identity over time.
However, in both cases the obvious temptation --- to modify the
formalism so as to solve the problem by \emph{fiat} --- has proved to be
unnecessary: it has been possible to find solutions within the existing
theory, and thus to preserve those features of the Everett
interpretation which made it attractive in the first place.

Something similar may be going on with the other major problem
of the Everett interpretation: that of understanding quantitative
probability. One of the most telling criticisms levelled at Everettians
by their critics has always been their inability to explain why, when
all outcomes objectively occur, we should regard one as more likely than
the other. The problem is not merely how such talk of probability can be
meaningful; it is also how the \emph{specific} probability rule used in
quantum mechanics is to be justified in an Everettian context. Here too, the temptation
is strong to modify the formalism so as to include the probability rule
as an explicit extra postulate.

Here too, it may not be necessary. David Deutsch
has, I believe, transformed the debate by
attempting \cite{deutschprobability} 
to derive the probability rule \emph{within} unitary quantum
mechanics, via considerations of rationality (formalised in 
decision-theoretic terms). His work has not so far met with wide
acceptance, perhaps in part because it does not make it at all obvious
that the Everett interpretation is central (and his proof manifestly
fails without that assumption). 

In \citeN{decshort}, I have presented an exegesis of Deutsch's proof in
which the Everettian assumptions are made explicit. The present paper
may be seen as complementary to my previous paper: it presents an
argument in the spirit of Deutsch's, but rather different in detail. My
reasons for this are two-fold: firstly, I hope to show that the
mathematics of Deutsch's proof can be substantially simplified, and his
decision-theoretic axioms greatly weakened; secondly and perhaps more
importantly, by simplifying the mathematical structure of the proof, I
hope to be able to give as clear as possible a discussion of the
\emph{conceptual} assumptions and processes involved in the proof. In
this way I hope that the reader may be in a better position to judge
whether or not the probability problem, like other problems before it,
can indeed be solved without modifications to the quantum formalism.

The structure of the paper is as follows. In section \ref{branching} I
review the account of branching that Everettians must give, and
distinguish two rather different viewpoints that are available to them;
in section \ref{weight} I consider how probability might
fit into such an account.  Sections \ref{axioms}--\ref{proof} are the
mathematical core of the paper: they present an extremely minimal
set of decision-theoretic assumptions and show how, in combination with an assumption which I
call \textbf{equivalence}, they are sufficient to derive the quantum
probability rule. The next three sections are a detailed examination of
this postulate of \textbf{equivalence}. I argue that it is
unacceptable as a principle of rationality for single-universe
interpretations (section \ref{single}), but is fully defensible for
Everettians --- either via the sort of arguments used by Deutsch
(section \ref{neutrality}) or directly (sections \ref{direct}--\ref{branchingindifference}). Section
\ref{conclusion} is the conclusion.

\section{Thinking about branching}\label{branching}

The conceptual problem posed by branching is essentially one of
transtemporal identity: given branching events in my future, how can it
even make sense for me to say things like 
``I will experience such-and-such''? Sure, the theory predicts that
people who look like me will have these experiences, but what
experiences will \emph{I} have? Absent some rule to specify \emph{which}
of these people is me, the only options seem to be (1) that I'm all of
them, in which case I will presumably have all of their experiences, or
(2) that I'm none of them, which seems to suggest that branching events
are fatal to me.

As \citeN{saundersprobability} has forcefully argued, this is a false
dilemma. Even in classical physics, it is a commonplace to suppose that
transtemporal identity claims, far from being in some sense primitive,
supervene on structural and causal relations between momentary regions
of spacetime. Furthermore, the work of \citeN{parfit} and others on fission,
teletransportation and the like has given us reason to doubt that
identity \emph{per se} is what is important to us: rather, it is the
survival of people who are appropriately (causally/structurally) related to 
me that is important, not \emph{my} survival \emph{per se}. In a
branching event, then, what is important is that the post-branching
people do indeed bear these relations to me. That I care about their
future well-being is no more mysterious than --- indeed, is precisely
analogous to --- my caring about my own future well-being.

If this resolves the paradox, still it leaves a choice of ways for
Everettians to think about splitting (which is the reason for this
section). The choice has not been discussed much in print (though see \citeNP{greaves} and
\citeNP{wallaceepist}), but 
it often arises in informal discussions and is of some relevance to
the rest of this paper. The two options are:
\begin{itemize}
\item The \emph{Subjective-uncertainty (SU) viewpoint:} Given that what
it is to have a future self is to be appropriately related to a certain
future person, and that in normal circumstances I expect to become my
future self, so also in Everettian splittings I should expect to become \emph{one} 
of my future selves. If there is more than one of them I should be uncertain as to which I will
become; furthermore, this subjective uncertainty is compatible with my \emph{total
knowledge} of the wavefunction and its dynamics. This is Saunders' own
view, argued for at length in \citeNP{saundersprobability}; 
I discuss his argument for SU in \citeN{wallaceepist}, and give my own defence of it in
\citeN{wallacebranching}.

(`Subjective' should not be taken too literally here. The subjectivity
lies in the essential role of a particular location in the quantum
universe (uncertainty isn't visible from a God's-eye view. But it need
not be linked to first-person expectations: `there will be a sea battle
tomorrow' might be as uncertain as `I will see spin up'.)
\item The \emph{Objective-determinism (OD) viewpoint:} 
Branching leads, deterministically, to my having multiple future descendants. 
Rationally speaking, I should act to benefit my future
descendants, for exactly the same reason that people in non-branching possible worlds
would act to benefit their single descendant. Situations of conflict may arise
between the interests of my descendants (such as when I bet on one possible outcome of a measurement), 
in which case I will have to weigh up how much I wish to prioritise each descendant's interests. 
This is perhaps the most literal translation of Parfit's
own ideas into a quantum-mechanical context; it appears to be the view
espoused by \citeN{bostrom} in his discussion of the Everett
interpretation, and is defended explicitly by \citeN{greaves}.
\end{itemize}
These views should not be taken as automatically in conflict.\footnote{Although I erroneously took
them as such in an earlier draft of this paper.} It is
almost impossible \emph{not} to accept the OD viewpoint as valid, since
it is just a literal reading of the physics. The conflict is rather
between those (such as Saunders and myself) who regard the SU viewpoint
as a valid alternative, and those (such as \citeN{greaves}) who regard
it as incoherent. There is a further question as to whether anything
important depends on the validity or otherwise of SU --- in my view it is
of central, albeit rather philosophical, importance to the epistemology of the Everett interpretation
(see \citeN{wallaceepist} for the argument) but others regard it as a
purely linguistic distinction.

One of the claims of this paper is that the Born rule can be defended
from both the OD and the SU viewpoints, albeit in slightly different
ways. I will return to these matters in sections \ref{neutrality} 
and \ref{branchingindifference}. For now,
however, let us move from the question of identity to the question of probability.

\section{Weight and probability}\label{weight}

The paradigm of a quantum measurement is something like this: prepare a
system (represented by a Hilbert space \mc{H}) in some state
(represented by a normalised vector $\ket{\psi}$ in \mc{H}). Carry out
some measurement process (represented by a discrete-spectrum self-adjoint operator
\op{X} over \mc{H}) on the system, and look to see what result
(represented by some element of the spectrum of \op{X}) is obtained.

Suppose some such measurement process is denoted by $M$, and
suppose that associated with $M$ is some set $\mc{S}_M$ of
possible outcomes of the process (for instance, states of the apparatus 
with pointers pointing in certain directions.) Let us call this the \emph{state
space} of the measurement process; and let us call $\mc{E}_M$, the set of
all subsets of $\mc{S}_M$, the \emph{event space} for $M$. We define the total event space for a set
of measurements as the union of all their event spaces.

If the observable being
measured is represented by operator \op{X}, then specifying the
measurement process requires us to specify some convention by which
elements of $\mc{S}_M$ are associated with elements of the spectrum
$\sigma(\op{X})$ of eigenvalues of \op{X}. In effect, the convention is a
function \mc{C} from $\mc{S}_M$ onto $\sigma(\op{X})$: $\mc{C}(s)$ is the outcome
which we associate with state $s$ (and $\mc{C}(E)$ is shorthand for
$\{\mc{C}(s)|s \in E\}$).

Now, suppose that the system being measured is in state \ket{\psi}. We
use it, and \mc{C}, to define a \emph{weight function} on $\mc{E}_M$, as follows:
\be \label{weightdef} \mc{W}_M (E) = \sum_{x \in
\mc{C}(E)}\matel{\psi}{\op{P}_X(x)}{\psi}\,\,\,\,\,\ \forall E \subseteq M,\ee
where $\op{P}_X(x)$ projects onto the eigenspace of \op{X} with
eigenvalue $x$.

All this should be both familiar and essentially interpretation-neutral;
familiar, too, should be the
\begin{quote}
\textbf{Quantum probability rule:} If $M$ is a quantum measurement and
$E \in \mc{E}_M$, then the probability of $E$ given that $M$ is
performed is equal to $\mc{W}_M(E)$.
\end{quote}

Familiar though it may be, do we actually understand it? Compare it with
the
\begin{quote}
\textbf{Quantum sqwerdleflunkicity rule:} If $M$ is a quantum measurement and
$E \in \mc{E}_M$, then the sqwerdleflunkicity of $E$ given that $M$ is
performed is equal to $\mc{W}_M(E)$.
\end{quote}

We \emph{don't} understand the quantum sqwerdleflunkicity rule, and for
an obvious reason: ``sqwerdleflunkicity'' is meaningless, or at any rate
we have no idea what it is supposed to mean. So if we \emph{do}
understand the quantum probability rule, presumably this requires that we
understand what ``probability'' means.

But in fact, the meaning of ``probability'' is pretty subtle, and pretty
controversial. In practice, physicists tend to test ``probability''
statements by relating them to observed frequencies, and as a result
most physicists tend to \emph{define} probability as relative frequency
in the limit. This is also the notion of probability used by most
investigators of the Everett interpretation, beginning with Everett's own analysis of
memory traces under repeated measurements and leading to
the elegant relative-frequency theorems established by \citeN{hartle} and
\citeN{FGG}. The general strategy of these investigations is to show
that as we approach the limiting case of infinitely many measurements, 
the weights of those branches in which relative frequencies are anomalous approach zero
(or alternatively that in mathematical models appropriate to an
\emph{actual} infinity of measurements, there are \emph{no} branches
with anomalous frequencies.)

Such arguments have in general failed to convince sceptics. Arguably
they either invoke unphysical situations such as an actual infinity of
experiments, or they court circularity: if we wish to \emph{prove} that the
quantum weight function is something to do with probability, we aren't
entitled to \emph{assume} that anomalous branches can be neglected just
because they have very low weight.

Now, anyone familiar with the chequered history of frequentist theories
of probability should feel uneasy at this point: virtually identical
criticisms could easily be levelled at the frequentist definition of
probability itself. For that definition, too, either makes use of
infinite ensembles or runs into the problem that anomalous frequencies
are just \emph{very improbable}, not actually impossible. It is most
unclear that Everettian frequentists are any worse off than other
species of frequentist. 

Nevertheless, if Everettians have an incoherent theory of probability 
it is cold comfort for them if other people do too. Is there a more
positive step that they can make? They could simply take probability as
primitive, and declare it to be an interpretative posit that
probability=weight; at one point this was  Simon Saunders' strategy
(see \citeN{saundersprobability}), and he has defended (successfully, in my view) the
position that non-Everettian theories of physics do no better. Again,
though, this seems unsatisfactory: we would like to \emph{understand}
Everettian probability, not just observe that non-Everettians also have
problems with probability.\footnote{Having said which, if \emph{no-one}
has a good theory of probability then it would be unreasonable to
dismiss the Everett interpretation in favour of other interpretations
purely on the grounds of the probability problem; see \citeN{papineau}
for a more detailed presentation of this argument.}

To the best of my knowledge, \citeN{deutschprobability} made the first concrete
non-frequentist proposal for how probabilities could be derived in the
Everett interpretation . His strategy follows in the
footsteps\footnote{I should note for the record that Deutsch himself dislikes this description of his
project (private conversation), essentially because of his deep skepticism about the coherence
of the subjective notion of probability. Nonetheless, at least from a
mathematical perspective the description is hard to challenge: Deutsch
appeals explicitly to the axioms of decision theory, as originated by
Savage et al.}
of the subjectivist tradition in foundations of probability,
originally developed by \citeN{ramsey}, \citeN{definetti}, \citeN{savage}
and others: instead of reducing probability to frequencies,
\emph{operationalise} it by reducing it to the preferences of rational
agents. We might then say that one such agent judges $E$ more likely
than $F$ iff that agent would prefer a bet on $E$ to a bet on $F$ (for the same stake); 
we
could similarly say that $E$ is more likely than $F$ \emph{simpliciter}
if \emph{all} agents are rationally compelled to prefer bets on $E$ to
bets on $F$. 

In principle, we might go further. An agent judges $E$ and $F$ to be
\emph{equally likely} iff he is indifferent between a bet on $E$ and one
on $F$; if he judges disjoint events $E$ and $F$ to be equally likely,
and judges $G$ to be equally likely as $E \cup F$, then we say that he
judges $G$ to be \emph{twice as likely} as $E$. By this sort of strategy, we can justify not
just qualitative but quantitative comparisons of likelihood, and --- perhaps --- ultimately
work up towards a numerical measure of likelihood: that is, a numerical probability.

The strategy of the subjectivists, then, was this: to state intuitively
reasonable axioms for rational preference, such that if an agent's
preferences conform to those axioms then they are provably required to be given by some probability
function. Deutsch essentially takes this strategy over to quantum
mechanics, but with a crucial difference: he uses the operationalist notion of probability not only to
\emph{make sense} of probabilistic talk within the Everett
interpretation, but also to \emph{prove} that the rational agents must use the weight
function $\mc{W}$ of equation (\ref{weightdef}) to determine probabilities. 

(It might appear from the above that probability in the Everett
interpretation is somehow ``not objective''. This is certainly not the
case: the weights of quantum branches are as objective as any other
physical property. In fact, the best reading of the decision-theoretic
proofs, in my view, is not that they tell us that there are no objective
probabilities, but rather that they teach us that objective probability
\emph{is} quantum weight. See \citeN{saundersquovadis} or
\cite{wallaceepist} for a more detailed analysis of this point.)

\section{A rudimentary decision theory}\label{axioms}

In this section I will develop some of the formal details of the subjectivist program, 
in a context which will allow ready application to quantum theory. 
Our starting point is the following: define a 
\emph{likelihood ordering}
as some two-place relation holding between ordered pairs $\langle E,M\rangle$, where 
$M$ is a quantum measurement and $E$ is an event in $\mc{E}_M$ (that is, $E$ is a subset of the
possible outcomes of the measurement). We write the relation 
as $\succeq$:
\be E|M \succeq F|N \ee
is then to be read as ``It's at least as likely that some outcome in $E$ will obtain 
(given that measurement $M$
is carried out) as it is that some outcome in $F$ will obtain (given that measurement $N$
is carried out)''. We define $\simeq$ and $\succ$ as follows: $E|M\simeq F|N$ if
$E|M\succeq F|N$ and $F|N\succeq E|M$; $E|M\succ F|N$ if
$E|M\succeq F|N$ but $E|M \simeq\crossout \,F|N$. We define $\preceq$
and $\prec$ in the obvious way, as the inverses of $\succeq$ and $\succ$
respectively. 

We will say that such an ordering is \emph{represented} by a function $\Pr$ from pairs $\langle E,M\rangle$
to the reals 
if
\begin{enumerate}
\item $\Pr(\emptyset|M)=0$, and $\Pr(\mc{S}_M|M)=1$, for each $M$.
\item If $E$ and $F$ are disjoint then $\Pr(E\cup F|M)=\Pr(E|M)+\Pr(F|M)$.
\item $\Pr(E|M)\geq\Pr(F|N)$ iff $E|M \succeq F|N$.
\end{enumerate}
The ordering is \emph{uniquely represented} iff there is only one such $\Pr$.

The subjectivist program then seeks to find axioms for $\succeq$ so that any agent's preferences
are uniquely represented.
Literally dozens of sets of such axioms have been
proposed over the years (see \citeN{fishburn} for a review). Their forms
vary widely, but as a rule there is an inverse correlation between the
complexity of the axioms and their individual plausibility. Deutsch, for
instance, uses a fairly simple but implausibly strong set of axioms
(which I reconstruct in \citeNP{decshort}).

However, in a \emph{quantum-mechanical} context we can manage with a set of axioms which
is both extremely weak --- far weaker than Deutsch's set --- and fairly simple.  To state them,
it  will be
convenient to  define a \emph{null event}: an event $E$ is 
\emph{null with respect to M} (or, equivalently, $E|M$ is null) iff $E|M \simeq \emptyset|M$. 
(That is: $E$
is certain not to happen, given $M$).
If it is clear which $M$ we're referring to, we will sometimes drop the $M$ and refer to $E$ as 
null \emph{simpliciter}.

We can then say
that a likelihood ordering is \emph{minimally rational} if it satisfies the
following axioms:
\begin{description}
\item[Transitivity] $\succeq$ is transitive: if $E|M\succeq F|N$ and $F|N \succeq G|O$, 
then $E|M\succeq G|O$.
\item[Separation]
There exists some $E$ and $M$ such that $E|M$ is not null.
\item[Dominance]
If $E \subseteq F$, then $F|M \succeq E|M$ for any $M$, with $F|M \simeq
E|M$ iff $E-F$ is null.
\end{description}

This is an extremely weak set of axioms for qualitative likelihood (far
weaker, for instance, than the standard de Finetti axioms --- see
\citeNP{definetti}). Each, translated into words, should be immediately
intuitive:
\begin{enumerate}
\item Transitivity: `If A is at least as likely than B and B is at least as likely
than C, then A is at least as likely than C.'
\item Separation: `There is some outcome that is not impossible.'
\item Dominance: `An event doesn't get less likely just because more
outcomes are added to it; it gets more likely iff the outcomes which are added are not themselves
certain not to happen.'
\end{enumerate}

\section{A quantum representation theorem}\label{proof}

It goes without saying that this set of axioms alone is insufficient to derive the quantum
probability rule: absolutely no connection has yet been made between the decision-theoretic
axioms and quantum theory. We can make this connection, however, via two further posits. Firstly, 
we need to assume that we have a fairly rich set of quantum measurements
available to us: rich, in fact, in the sense of the following
definition.

\begin{quote}
\textbf{Weight richness:} A set \mc{M} of quantum measurements is \emph{rich} 
provided that, for any positive real numbers $w_1, \ldots w_n$ with $\sum_{i=1}^n=1$, \mc{M} 
includes a quantum measurement with $n$ outcomes having weights
$w_1, \ldots, w_n$.
\end{quote}

The richness of a set of measurements is an easy consequence of the following
(slightly informal) principle: that for any $n$ there exists
at least one system whose Hilbert space has dimension $n$ such that that
system can be prepared in any state and that at least one 
non-degenerate observable can be measured on that system. Clearly, this
is an idealisation: 
with sufficiently high computational power we can prepare
states with arbitrary accuracy, but presumably there is some limit to
that power in a finite universe. However, it seems a reasonable 
idealisation (just as it is reasonable to idealise the theory of
computations slightly, abstracting away the fact that in practice the
finitude of the universe puts an upper limit on a computer's memory)
and I will assume it without further discussion.

Far more contentious is the second principle which we must assume:
\begin{description}
\item[Equivalence]
If $E$ and $F$ are  events and $\mc{W}_M(E)=\mc{W}_N(F)$, then $E|M \simeq F|N$.
\end{description}
Much of the rest of the paper will be devoted to an analysis of whether
\textbf{equivalence} is a legitimate requirement for rational agents.
But the point of such an analysis, of course, is that in the 
decision-theoretic context which we are analysing, a great deal can be
proved from it. In fact, we are now in a position to prove the
\begin{description}
\item[Quantum Representation Theorem:]
Suppose that $\succeq$ is a minimally rational likelihood order for a rich set of
quantum measurements, and suppose that $\succeq$ satisfies \textbf{equivalence}.
Then $\succeq$ is uniquely represented by the
probability measure $\Pr(E|M)=\mc{W}_M(E)$.
\end{description}

\noindent\textbf{Proof:}
We proceed via a series of lemmas.
\begin{lemma}
If $\mc{W}_M(E)\geq \mc{W}_N(F)$, then $E|M \succeq F|N$. 
\end{lemma}
Since the set of quantum measurements is rich, there exists a quantum measurement $O$ with disjoint events
$G,H$ such that $\mc{W}_O(G)=\mc{W}_N(F)$ and $\mc{W}_O(H)=\mc{W}_M(E)-\mc{W}_N(F)$. 
By \textbf{equivalence} $E|M \simeq G \cup H_O$ and $G|O \simeq F|N$; by \textbf{dominance}
$G \cup H|O \succeq G|O$; by \textbf{transitivity} it then follows that
$E|M \succeq F|N$.

\begin{lemma}
A quantum event is null iff it has weight zero. 
\end{lemma}
That weight-zero events are null follows from 
\textbf{equivalence} and $\mc{W}_M(\emptyset)=0$. 
For the converse, suppose for contradiction that some event of weight $w>0$ 
is null; then there must exist $n$ such that $1/n<w$. By lemma 1, any event of weight $1/n$ must also
be null. Since the set of measurements is rich, there exists a quantum measurement $M$ 
with $n$ outcomes all of weight $1/n$: all must be null,
and so by repeated use of \textbf{dominance} so must $\mc{S}_M$.
By lemma 1 we conclude that all events are null, in 
contradiction with \textbf{separation}.

\begin{lemma}
If $E|M \succeq F|N$, then $\mc{W}_M(E)\geq \mc{W}_N(F).$
\end{lemma}
Suppose that  $\mc{W}_M(E)< \mc{W}_N(F)$. The set of measurements is rich, so there exists
a quantum measurement $O$ with disjoint events $G,H$ such that $\mc{W}_O(G)=\mc{W}_M(E)$ and
$\mc{W}_O(H)=\mc{W}_N(F)-\mc{W}_M(E)$. If $E|M \succeq F|N$ then $G|O \succeq G\cup H|O$, which
by \textbf{dominance} is possible only if $H$ is null.
But since $\mc{W}_O(H)>0$, by lemma 2 $H\simeq\crossout\,\,\emptyset$.
\\

Lemmas 1 and 3 jointly prove that $\mc{W}$ represents $\succeq$. 
To show that the representation is
unique, suppose $M$ is any quantum measurement with outcomes having 
weights $k_1/K,k_2/K,\ldots k_N/K$, where $k_1$ through $k_n$ are positive integers whose
sum is $K$. Since the set of measurements is rich, there exists
another measurement $M'$ with $K$ possible outcomes each having weight
$1/K$. Since each is equiprobable, any probability function representing $\succeq$
must assign probability $1/K$ to each outcome. Let $E_1$ be the union of
the first $k_1$ outcomes of $M'$, $E_2$ be the union of the next $k_2$
outcomes, and so on; then any probability function must assign
probability $k_i/K$ to $E_i|M'$, and hence (since they have equal weight) 
to the $i$th outcome of $M$.

So: any probability function must agree with the weight function on
rational-weight events. But since any irrational-weight event is
more likely than all rational weight events with lower weight
and less likely than all rational weight events with higher
weight, agreement on rational values is enough to force agreement on all
values.$\Box$

\section{Equivalence and the single universe}\label{single}

So far, so good. The ``decision-theoretic turn'' suggested by Deutsch
not only allows us to \emph{make sense} of probability in an Everettian
universe, but it also allows us to derive the quantitative form of the
probability rule from assumptions --- \textbf{equivalence} and the
richness of the set of measurements --- which \emph{prima facie} are
substantially weaker than the rule itself.

Nonetheless the situation remains unsatisfactory. \textbf{Equivalence}
has the form of a principle of pure rationality: it dictates that any
agent who does not regard equally-weighted events as equally likely is
in some sense being irrational. Whether this principle is simple,  
or ``\emph{prima facie} weak'',
is not the point: the point is whether it is defensible purely from
considerations of rationality. Simplicity might be a virtue
when we are considering which axioms of fundamental physics to adopt,
but can our ``axioms of fundamental physics'' include statements which
speak directly of rationality? Surely not.

In fact, I believe that \textbf{equivalence} can be defended on grounds
of pure rationality, and need not be regarded as a new physical axiom; I
shall spend sections \ref{neutrality}--\ref{branchingindifference} providing such a defence. However,
before doing so I wish to argue that the Everett interpretation necessarily plays 
a central role in any such defence: in other interpretations, \textbf{equivalence} is
not only unmotivated as a rationality principle but is actually
absurd.\footnote{In this section I confine my observations to those
interpretations of quantum mechanics which are in some sense ``realist''
and observer-independent (such as collapse theories or hidden-variable
theories). I will not consider interpretations (such as the Copenhagen
interpretation, or the recent variant defended by \citeNP{fuchsperes})
which take a more `operationalist' approach to the quantum formalism. It
is entirely possible, as has been argued recently by
\citeN{saundersoperation}, that an approach based on Deutsch's proof may
be useful in these interpretations.}

Why? Observe what \textbf{equivalence} actually claims: that if we know
that two events have the same weight, then we must regard them as
equally likely \emph{regardless of any other information we may have
about them}. Put another way, if we wish to determine which event to bet
on  and we are told that they have the same weight, we will be
uninterested in any other information about them.

But in any interpretation which does not involve branching --- that is,
in any non-Everettian interpretation --- there is a further
piece of information that cannot but be relevant to our choice: namely,
\emph{which event is actually going to happen}? If in fact we know that
$E$ rather than $F$ will actually occur, \emph{of course} we will bet on
$E$, regardless of the relative weights of the events. 

This objection can be made precise in one of two ways. The first might
be called the argument from \emph{fatalism}, and presumes a B-theoretic
(or indexical, or `block-universe') view of time: that from God's
perspective there is no difference between past, present and future. If
this is the case, then there is a \emph{fact of the matter} (regardless
of whether the world is deterministic) as to which future event occurs.
Maybe this fact is epistemically inaccessible to us even in principle;
nonetheless \emph{if} we knew it, of course it would influence our bets.

Maybe you don't like the B-theoretic view of time; or maybe you see
something objectionable in appeal to in-principle-inaccessible facts.
Then I commend to your attention the second way of making the objection
precise: the argument from \emph{determinism}. Suppose, for the moment,
that quantum theory is deterministic and non-branching: for instance, it
might be the de Broglie-Bohm pilot-wave theory, or some other
deterministic hidden-variables theory). Then a sufficiently complete
description of the microstate will determine exactly which event will
occur, regardless of weights; again, if we had this complete
description it would certainly influence (indeed, fix) our preferences
between bets.

(Maybe certain details of the microstate are ``in principle inaccessible'', 
as is arguably the case in the pilot-wave theory. This doesn't improve
matters: for the theory to have any predictive power at all we need to
say \emph{something} about the hidden variables, and in practice we need
to give a probability distribution over them. Well, ``hidden variables
are randomly distributed in such-and-such a way'' might be a reasonable 
\emph{law of physics}, but absent such a
law, there seems no justification at all for adopting ``it is rational
to assume such-and-such distribution of hidden variables''.)

What if quantum theory is \emph{in}deterministic? Again, if the theory
is to have predictive power then we must replace deterministic evolution
with something else: a stochastic dynamics.
In this case, it only makes
sense to adopt \textbf{equivalence} if the stochastic dynamics makes
equal-weighted events equiprobable --- and as Barnum \emph{et al} \citeyear{barnumetal} have
pointed out in their discussion of Deutsch's work, there is no difficulty in 
constructing a stochastic dynamics for quantum mechanics
which does no such thing.

The reason that the Everett interpretation is not troubled by these
problems is simple. Regardless of our theories of time, and
notwithstanding the determinism of her theory, for the Everettian 
there is simply \emph{no fact of the matter} which measurement outcome
will occur, and so it is not just impossible, but incoherent for an
agent to know this fact. This coexistence of determinism with the 
in-principle-unknowability of the future is from a philosophical point
of view perhaps the Everett interpretation's most intriguing feature.

\section{Equivalence via measurement neutrality}\label{neutrality}

If we have shown that \textbf{equivalence} is implausible except in the Everett
interpretation, still we have not shown that it is plausible for
Everettians; that is our next task. An obvious starting point is
Deutsch's original work; but Deutsch makes no direct use of
\textbf{equivalence}. Instead, he uses --- implicitly, but extensively 
--- a principle which I have elsewhere \cite{decshort} called
\textbf{measurement neutrality}: the principle that once we have
specified which system is being measured, which state that system is
being prepared in, and which observable is being measured on it, then we
have specified everything that we need to know for decision-making
purposes. In this section I will show how \textbf{measurement
neutrality} is sufficient to establish \textbf{equivalence}, and discuss
whether \textbf{measurement neutrality} is itself justifiable; in the
next section I will look at more direct arguments for
\textbf{equivalence}.

In effect, assuming \textbf{measurement neutrality} allows us to
replace the abstract set $\mc{M}$ of measurements with the set of
triples $\langle \mc{H},\ket{\psi},\op{X}\rangle$, and the abstract set 
$\mc{S}_M$ of outcomes of the measurement with the spectrum
$\sigma(\op{X})$ of \op{X}.
The relation $\succeq$ is similarly transferred from ordered pairs $\langle E,M\rangle$
to ordered quadruples $\langle E,
\mc{H},\ket{\psi},\op{X}\rangle$ (where $E \subseteq
\sigma(\op{X})$ is now just a subset of $\Re$).

From a purely mathematical point of view, this move clearly gets us no
closer to deriving \textbf{equivalence} (and, thus, the quantum
representation theorem), and consequently most commentators on Deutsch's
work (\eg Barnum \emph{et al} \citeyearNP{barnumetal}; \citeNP{gill,peterlewis}) have claimed that he is guilty of a 
\emph{non sequitur} at various
points in his derivation. In fact (as I argued in \citeNP{decshort}) he
is guilty of no such thing.

The central insight of Deutsch's argument is that when measurement
processes are represented physically there can be no unambiguous way to
assign triples $\langle \mc{H},\ket{\psi},\op{X}\rangle$ to these
physical processes. Hence, one and the same process may be validly
represented by two triples --- and if so, of course, the relation
$\succeq$ should not distinguish between those triples.

How does the ambiguity arise? For one thing,
the physical process of preparing a state and then measuring
an observable on it consists of a large number of unitary
transformations applied to the state and to various auxiliary systems
and there is no privileged moment at which the preparation phase can be
said to have finished and the measurement phase begun; nor is there any privileged way of saying which
system is the one being measured and which is the `auxiliary' system. 

Consider, for instance, the Stern-Gerlach experiment (see, for instance,
\citeN{feynmanlecture} for an elementary discussion). A beam of particles prepared in some
spin state is placed in an inhomogeneous magnetic field in the $+z$ direction, with the result that particles
with spin up (in the $z$ direction) are deflected in one direction and particles of spin down
in another. The outgoing beams are incident on some detector which
records the location of each particle. 

The Stern-Gerlach device is generally treated as a paradigmatic
measurement of particle spin --- in this case along the $z$ axis. Now
suppose that we wish to apply that measurement to particles in  spin
state $\ket{+_x}=\frac{1}{\sqrt{2}}(\ket{+_z}+\ket{-_z})$, but our preparation device
only outputs spin state $\ket{+_z}$. There is an easy remedy: after the
particles emerge from the preparation device, expose them to a homogenous magnetic
field which causes their spins to precess, taking $\ket{+_z}$ to the
desired state $\ket{+_x}$. Effectively, the new magnetic field is just
an extra part of the preparation device.

Conversely, suppose that we want to measure the spin along the $-x$ axis
but that it is too difficult physically to rotate the Stern-Gerlach
apparatus. Again there is a simple solution: instead of rotating the apparatus, rotate
the particles by exposing them to the same magnetic field. That magnetic
field is effectively just part of the measurement device.

However, from a purely physical viewpoint there is no difference at all
between these two processes. In each case, we prepare particles in state
\ket{+_z}, expose them to a magnetic field, and then insert them into a
Stern-Gerlach device aligned along the $+z$ axis. In the one case we
have regarded the magnetic field as part of the preparation, in the
other case as part of the measurement --- but since this difference
is purely a matter of convention and does not correspond to any
\emph{physical} difference, 
it should be regarded as irrelevant to the subjective likelihood of detecting
given results.

In fact, we can go further than this: it is also only a matter of
convention that we take the process to be a measurement of spin and not
of position. For although we have described the splitting of the beam
into two as part of the measurement process, it could equally well be
regarded as part of a state preparation --- in this case, to prepare a 
particle in a coherent superposition of two
positions.\footnote{Technically the particles are actually in an
entangled state, since their spin remains correlated with their
position; the important point remains that the observable being measured
could be taken to be a joint spin-position observable, rather than a
pure spin observable. (In any case, removing this entanglement in the
Stern-Gerlach measurement process would be technically possible, albeit
rather difficult in practice.)}

The moral of this example is as follows: any process which we describe
as `prepare a state, then measure it' actually consists of a long
sequence of unitary transformations on a wide variety of quantum systems
(for instance, the magnet performs a unitary transformation on the spin
space of the particle, and then the Stern-Gerlach apparatus performs
another on the joint spin-position space of the particle). As it is
purely conventional when these transformations stop being part of the
state preparation and start being part of the measurement process,
probabilistic statements about that process cannot depend on that
convention. Symbolically, this is to say that 
there will be some processes which are equally well
described by $\langle\mc{H},\ket{\psi},\op{X}\rangle$ and by 
$\langle\mc{H}',\op{U}\ket{\psi},\op{X}'\rangle$ where
$\op{U}:\mc{H}\rightarrow\mc{H}'$ is a unitary transformation and
$\op{U}\op{X}=\op{X}'\op{U}$. We can
write this as
\be \label{simone} \langle E,\mc{H},\ket{\psi},\op{X}\rangle \sim
\langle E, \mc{H}',\ket{\psi'},\op{X}'\rangle,\ee
which is to be read as
`there is some physical process represented both as a measurement of
\op{X} on \ket{\psi} followed by a bet on $E$, and as 
a measurement of
$\op{X}'$ on \ket{\psi'} followed by a bet on $E$.'

Similarly, the physical process of reading off the result of a
measurement from the apparatus involves both the physical state of the
apparatus \emph{and} some convention as to which number is associated
with which physical state --- even if the device has a needle pointing
to the symbol ``\textbf{6}'' then it is a human convention and not a law
of physics that associates that symbol with the sixth positive integer.
But if so, then it is purely a matter of convention whether a
measurement of \op{X} which obtained result $x$ should actually be
regarded as a measurement of $f(\op{X})$ which obtained result $f(x)$. 

Again, conventions should not affect our judgements about likelihood, if
they
do not correspond to anything physical. As such, it follows that 
\be \label{simtwo}\langle E,\mc{H},\ket{\psi},\op{X} \rangle\sim
\langle f(E),\mc{H},\ket{\psi},f(\op{X})\rangle\ee 
for arbitrary $f$. 

From (\ref{simone}) and (\ref{simtwo}) we can establish
\textbf{equivalence}. For if we take $f(x)=0$ whenever $x\in E$ and
$f(x)=1$ otherwise, it follows from (\ref{simtwo}) it follows that if $M$ is represented by the 
quadruple $\langle E,\mc{H},\ket{\psi},\op{X}\rangle$ then that quadruple is $\sim$-equivalent to one
of form $\langle \{0\},\mc{H},\ket{\psi},\op{X}_0,\rangle$, where
$\op{X}_0$ has only $0$ and $1$ as eigenvalues.

Now let $\mc{H}_0$ be a fixed two-dimensional Hilbert space spanned by vectors $\ket{0}$
and $\ket{1}$. There will exist some unitary operator \op{U} from $\mc{H}_0$ to
$\mc{H}$ which takes $\ket{0}$ to the $0$-eigenspace of \op{X} and
$\ket{0}$ to the $1-$eigenspace of \op{X}. This operator satisfies
$\op{U}\proj{1}{1} = \op{X}\op{U}$; 
from (\ref{simone}) it then follows 
that $\langle \{0\},\mc{H},\ket{\psi},\op{X}_0,\rangle$ is $\sim$-equivalent to
\be\langle\{0\},\mc{H}_0,c\ket{0}+d\ket{1},\proj{1}{1}\rangle,\ee
where $c^2=\mc{W}_M(E)$ and $c,d>0$. So 
$\sim$-equivalence classes are characterised entirely by the weights they give
to events.\footnote{For a slightly more detailed version of this
argument (in a mildly different notation) see \citeN{decshort}.}

So: \textbf{measurement neutrality} is sufficient to establish the quantum probability theorem,
given our physical definition of measurement.\footnote{It is perhaps worth noting that
Deutsch himself does not derive \textbf{equivalence} directly from \textbf{measurement neutrality},
but rather uses the latter to derive a number of special cases of \textbf{equivalence} which are
sufficient to prove his own form of the quantum representation theorem. See \citeN{decshort} for 
more details.} Furthermore, the
assumption is innocuous on its face:
it is part of the basic structure of quantum mechanics that
measurement processes are specified up to irrelevant details once we know what observable and 
state are to be measured.

Unfortunately, that innocuousness is somewhat misleading. The reasons why we treat 
the state/observable description as complete are not independent of the
quantum probability rule. On the contrary, a standard argument might go:
``the probability of any given outcome from a measurement process is
specified completely by state and observable, so two different systems
each described by the same state/observable pair will give the same
statistics on measurement.'' Of course, such a justification is in
danger of circularity in the present context.

In fact, it is possible to argue that from an Everettian (or indeed a
collapse-theoretic) viewpoint, ``measurements'' are a misnamed and even
unnatural category of processes. For traditionally the point of
``measurement'' is to learn something, whereas all that happens when we
``measure'' an already-known state in Everettian quantum mechanics is that we induce a
certain decoherent process (and all that happens in collapse-theoretic
quantum mechanics is that we trigger the collapse mechanism). In the
betting scenarios that are the topic of this paper, in fact,  there is in
a sense nothing to learn: \emph{ex hypothesi} the state is already known with certainty.

If `measurements' are an unnatural category, \textbf{measurement neutrality} seems
to be unmotivated: it's only decision-theoretically relevant that two processes fall under the same
description if we have reason to believe that that description captures
everything decision-theoretically relevant. As such, 
it is probably advisable for us to abandon
\textbf{measurement neutrality} altogether, and look for a more direct
justification of \textbf{equivalence}; that will be my strategy
in the next section. 

However, I remain unconvinced that
\textbf{measurement neutrality} is so unmotivated even for an
Everettian. For one thing, something seems wrong about the previous
paragraph's attack on `measurements'. It certainly fails to be an
accurate description of real physicists' \emph{modus operandi}: the
experimentalist building (say) a cloud chamber believes himself designing a device
that will detect particles, not one which induces decoherence, and the
design principles he employs are selected accordingly.
 
This is, I think, one point where the OD and SU viewpoints on branching
(discussed in section \ref{branching}) lead to different results. 
Someone who regards the SU viewpoint as incoherent is already committed
to the falsehood of much of our pre-theoretic discourse about quantum
mechanics; as such they will probably be prepared to bite the bullet and say that `real
physicists' are profoundly mistaken about what `measurement devices' are
(just as all of us are profoundly mistaken in regarding measurement
results as uncertain.) Without the SU viewpoint, I think, 
\textbf{measurement neutrality} has no available defence and it is
necessary to look directly for justifications of \textbf{equivalence}. (This is the conclusion drawn by \citeN{greaves}.)

Is it defensible for those who accept the SU viewpoint? As I discuss in
\citeNP{decshort}, they will agree with Dirac and von Neumann that a
measurement device performs two functions: it induces wave-function
collapse of the state being measured into some eigenstate of the
observable being measured, and it then evolves deterministically into a
state indicating the eigenvalue of that state. But, \emph{unlike} Dirac
and von Neumann, they will not find this dual function mysterious. For
an Everettian, the wave-function collapse is only ``effective'' and ``phenomenological''. 
More precisely, the branching structure of the universe (which in turn defines 
``effective collapse'') is defined
pragmatically, in terms of which structure(s) are explanatorily and
predictively most useful (I defend this view in
\citeNP{wallacestructure}). The very fact that the measurement process is
going to occur, and that it will rapidly lead to decoherence between the
various output states of the measurement device, guarantees that the
explanatorily and predictively natural branching structure to choose is one in which each branch finds
the system being measured in an eigenstate.

Thus, from the SU viewpoint there \emph{seems} (I go no further than
this) to be an important sense in which ``measuring devices'' really do
deserve that name, and hence in which \textbf{measurement neutrality} is
indeed innocuous.

\section{Equivalence, directly}\label{direct}

Perhaps the reader is not prepared to accept the SU viewpoint; or perhaps s/he
is unconvinced by the SU-dependent defence of \textbf{measurement
neutrality} which I offered above. Either way, it seems worth looking
directly at \textbf{equivalence}, to see if it can be justified without
recourse to \textbf{measurement neutrality}. This is also of interest
because it allows a direct reply to those critics of Deutsch (such as
\citeNP{peterlewis}) who argue that there can be no decision-theoretic reason
to be indifferent between choices that lead to very different branching
structures, even if they give the same quantum-mechanical weight to a
given outcome.

In fact, a very simple and direct justification of \textbf{equivalence}
is available. Consider, for simplicity, a Stern-Gerlach experiment of
the sort discussed in section \ref{neutrality}: an atom is prepared in a
superposition $\ket{+_x}=\frac{1}{\sqrt{2}}(\ket{+_z}+\ket{-_z})$ and
then measured along the $z$ axis. Accordiing to the result of the
measurement, an agent receives some payoff. \emph{Ex hypothesi} the
agent is indifferent \emph{per se} to what goes on during the
measurement process and to what the actual outcome of the experiment is;
all he cares about is the payoff.

We now consider two possible games (that is, associations of payoffs
with outcomes):
\begin{description}
\item[Game 1:] The agent receives the payoff iff the result is spin up.
\item[Game 2:] The agent receives the payoff iff the result is spin
down.
\end{description}
In each game, the weight of the branch where the agent receives the
payoff is 0.5; \textbf{equivalence}, in this context, is then the claim that the agent is
indifferent between games 1 and 2.

To see that this is indeed the case, we need to model the games
explicitly. Let $\ket{\mbox{`up';reward}}$ and $\ket{\mbox{`down'; no reward}}$ be the quantum states of the two
branches on the assumption that game 1 was played: that is, let the
post-game global state\footnote{More precisely: the global state relative to the pre-game agent: there
are of course all manner of other branches which are already effectively disconnected from the agent's branch.} if game 1 is played be
\be
\ket{\psi_1}=\frac{1}{\sqrt{2}}(\ket{\mbox{`up';reward}}+ \\ \ket{\mbox{`down'; no reward}}).\ee
Similarly, if game 2 is played then the quantum state is
\be
\ket{\psi_1}=\frac{1}{\sqrt{2}}(\ket{\mbox{`down';reward}}+ \\ \ket{\mbox{`up'; no reward}}).\ee
Why should an agent be indifferent between a physical process which
produces \ket{\psi_1} and one which produces \ket{\psi_2}?

Well, recall that the agent is indifferent \emph{per se} to the result
of the experiment. This being the case, he will not object if we erase
that result. Let \ket{\mbox{`erased',reward}} indicate the state of the
branch in which the reward was given  post-erasure and \ket{\mbox{`erased',no reward}}
the post-erasure state of the no-reward branch. Then (game 1+erasure)
leads to the state
\be
\ket{\psi_{1;e}}=\frac{1}{\sqrt{2}}(\ket{\mbox{`erased',reward}}+\ket{\mbox{`erased',no
reward}})\ee
and (game 2+erasure) to the state
\be
\ket{\psi_{2;e}}=\frac{1}{\sqrt{2}}(\ket{\mbox{`erased',reward}}+\ket{\mbox{`erased',no
reward}})\ee
--- that is, (game 1+erasure) and (game 2+erasure) lead to the same
state. If (game 1+erasure) and (game 2+erasure) are just different ways
of producing the same physical state --- different ways, moreover, which
can be made to differ only over a period of a fraction of a second, in
which the agent has no interest --- then the agent should be indifferent
between the two. Since he is also indifferent to erasure, he is
indifferent between games 1 and 2, as required by \textbf{equivalence}.

Actually, as it stands this is too quick. It is implausible that the
erasure process will lead to \emph{precisely} the same state in both
cases; in fact, microscopic differences (which the agent is
unaware of and indifferent to) are bound to persist, on pain of a failure of decoherence.
But this problem can be rectified as follows. Let
$\ket{\mbox{`erased(i)', reward}}$ denote one of the vastly many possible
quantum states which could be reached by erasure. Each of the
$\ket{\mbox{`erased(i)', reward}}$
is indistinguisable to the agent; furthermore, since he is indifferent
to the erasure (let alone to its details) he does not care which
$\ket{\mbox{`erased(i)', reward}}$
results from the erasure process. Let $\ket{\mbox{`erased(j)', no
reward}}$ have its obvious meaning.

The global state following (game 1+erasure) is then some
\be
\ket{\psi_1;i,j}=\ket{\mbox{`erased(i)', reward}}+\ket{\mbox{`erased(j)', no
reward}}
\ee 
and the agent does not care which one; that is, he is indifferent
between processes which produce any state in
\be
\mc{S}=\{\frac{1}{\sqrt{2}}(\ket{\mbox{`erased(i)', reward}}+\ket{\mbox{`erased(j)', no
reward}})|i,j\},
\ee
and furthermore he is indifferent between \ket{\psi_1} and any element
of \mc{S}. But of course, exactly the same argument tells us that
he is indifferent between \ket{\psi_2} and any element of \mc{S}; hence
that he is indifferent between \ket{\psi_1} and \ket{\psi_2}; hence that
he is indifferent between games 1 and 2.

(This argument actually gives some insight into why an agent should care
about the quantum weight. For suppose that we use an \emph{un}equal
superposition:
\be\ket{p}=\sqrt{p}\ket{+_z}+\sqrt{1-p}\ket{-_z}.\ee
The result of (game 1+erasure) will be some element of
\be
\mc{S}_p=\{\sqrt{p}\ket{\mbox{`erased(i)', reward}}+\sqrt{1-p}\ket{\mbox{`erased(j)', no
reward}}|i,j\};\ee
the result of (game 2+erasure) will be some element of
\be
\mc{S}_{1-p}=\{\sqrt{1-p}\ket{\mbox{`erased(i)', reward}}+\sqrt{p}\ket{\mbox{`erased(j)', no
reward}}|i,j\}.\ee
$\mc{S}_{1-p}=\mc{S}_p$ only when $p=0.5$.)

The generalisation to other weights is straightforward: just add a third
possible state (\ket{0_z}, say) of the system being measured, and define
games 1 and 2 as before. If the system is prepared in state
\be\sqrt{w}\ket{+_z}+\sqrt{w}\ket{-_z}+\sqrt{1-2w}\ket{0_z}\ee
then the global state after (game x+erasure) is some
\be
\ket{\psi;i,j,k}=\ket{\mbox{`erased(i)', reward}}+\ket{\mbox{`erased(j)', no
reward}}+\ket{\mbox{`erased(k)', no
reward}}\ee
whether we are playing game 1 or game 2. As for phase, this can be
incorporated by allowing phase changes in the erasure process: if
\ket{\mbox{`erased(i)', reward}} is a valid erasure state, so is 
$\exp(i \theta)\ket{\mbox{`erased(i)', reward}}$. More directly, it can be
incorporated by observing that a phase transformation of an entire branch is  
completely unobservable, so an agent should be indifferent to it.

Finally, recall that \textbf{equivalence} must hold even when the two
weight-equivalent rewards occur in different chance setups. This too can
be handled via erasure: simply erase all details of which particular
setup is under consideration, except for the weights of the payoff and
non-payoff branches.

\section{Branching indifference}\label{branchingindifference}

There are two closely related lacunae in the erasure proof of
\textbf{equivalence}. Firstly, erasure may lead to branching: realistic
erasure processes will usually lead not to a single \ket{\mbox{`erased(i)', reward}}
but to a superposition of them. Secondly, \textbf{equivalence} must hold
not just when we have two equally weighted branches, but when we have
one branch whose weight equals the combined weights of several other
branches.

Both of these lacunae would be resolved if we could establish
\begin{description}
\item[Branching indifference:] An agent is rationally compelled to be
indifferent about processes whose only consequence is to cause the world
to branch, with no rewards or punishments being given to any of his
descendants.
\end{description}

This principle is not at all obvious: why should I not care about
whether there is one of me or a thousand ten minutes from now? More generally, why should
the `branching microstructure' of constantly dividing worlds which underlies the macroscopic
ascription of weights to coarse-grained outcomes be decision-theoretically irrelvant? But it is
in fact irrelevant, for two distinct reasons; my last task in
this paper will be to establish this.

The
first argument for branching indifference works only from the SU viewpoint. 
From that viewpoint, recall,
branching events are to be understood 
as cases of \emph{subjective uncertainty}: the agent should expect to
experience one or other outcome, but does not (and cannot) know which.

But in this case, it is easy to see that the agent should be indifferent
to branching \emph{per se}. Suppose that someone proposes to increase a
million-fold the number of the agent's descendants who see heads: say,
by hiding within the measurement device a randomizer that generates and
displays a number from one to 1 million, but whose output the agent
doesn't care about and probably never sees. Then from the SU viewpoint,
this just corresponds to introducing some completely irrelevant extra
uncertainty. For it is the central premise of the SU viewpoint that
process which from an objective standpoint involves branching, may be
described subjectively as simply one with uncertain outcomes. In this
case the objective description is ``the agent branches into a million
copies who see heads, and one copy who sees tails''; the correct
description \emph{for the agent himself} is ``I will either see heads or
tails, and I am uncertain as to which; if I see heads then I am further
uncertain about the result of the randomiser reading --- but I don't
care about that reading''. 

But it is a (trivially) provable result of decision theory that
introducing ``irrelevant'' uncertainty of this kind is indeed irrelevant
(it is essentially the statement that if we divide one possible outcome
into equally-valuable suboutcomes, that division is not 
decision-theoretically relevant). As such, from the SU viewpoint
branching indifference follows trivially.

(Of course, a critic may deny that the observer's description really is
correct --- but this is simply another way to reject the SU viewpoint
itself.)

Everettians who reject the SU viewpoint cannot resort to this strategy, but
all Everettians can resort to the other argument: that it is not in fact
possible to pursue a non-branch-indifferent strategy in a quantum
universe. In part this is due to the fact that branching is going on all
the time, which leads to two objections to any proposed violation of
branch indifference:
\begin{enumerate}
\item The epistemic objection: to take decisions in such a universe, an
agent who was not branch indifferent would have to be keeping
microscopically detailed track of all manner of branch-inducing events
(such as quantum decays) despite the fact
that none of these events have any detectable effect on him. This is
beyond the plausible capabilities of any agent.
\item The small-world objection: it has long been recognised (see, \egc,
\citeNP{savage}) that decision-making will be impossibly complicated
unless it is possible to identify (in at least a rough-and-ready manner)
a point after which the dust has settled and the value to an agent of
consequences can actually be assessed. But if an agent is not branch
indifferent, then such a point will never occur, and he will be faced
with the impossible task of calculating how much branching will occur
across the entire lifetime of the Universe (contingent on his choice of
action) in order to weigh up the value, now, to him of carrying out a
certain act.
\end{enumerate}

Both of these objections rely on the assumption that a rational strategy
must actually be realisable in at least some idealised sense. In
earlier versions of this work  I took this as self-evident, but to my
surprise this has not generally been accepted (mostly this has emerged
in conversation; however, see also \citeNP{peterlewis}). I therefore offer
a brief defence:
\begin{enumerate}
\item Decision theory is something of a hybrid. It is to some extent
normative (that is, it tells us what we \emph{should} do, and exposes us
to rational criticism if we violate its precepts); it is to some extent
descriptive (that is, it provides an idealised account of actual
decision-making). Both of these are impossible if decision theory
instructs us to do something wildly beyond even our idealised abilities.
\item If we are prepared  to
be even slightly instrumentalist in our criteria for belief ascription,
it may not even \emph{make sense} to suppose that an agent genuinely
wants to do something that is ridiculously beyond even their idealised
capabilities. For instance, suppose I say
that I desire (\emph{ceteris paribus}) to date someone with a prime
number of atoms in their body. It is not even remotely possible for me
to take any action which even slightly moves me towards that goal. In
practice my actual dating strategy will have to fall back on
``secondary'' principles which have no connection at all to my
``primary'' goal --- and since those secondary principles are actually
what underwrites my entire dating behaviour, arguably it makes more
sense to say that \emph{they} are my actual desires, and that my
`primary' desire is at best an impossible dream, at worst an empty
utterance.
\end{enumerate}
However, even if this is not persuasive, then there is a stronger reason
why non-branch indifferent strategies cannot be pursued. Namely: 
non-branch-indifferent strategies require us to know the number of
branches, and there is no such thing.

Why? Because the models of splitting often considered in discussions of
Everett --- usually involving two or three discrete splitting events,
each producing in turn a smallish number of branches --- bear little or
no resemblance to the true complexity of realistic, macroscopic quantum
systems. In reality:
\begin{itemize}
\item Realistic models of macroscopic systems are
invariably infinite-dimensional, ruling out any possibility of counting
the number of discrete descendants. 
\item In such models the
decoherence basis is usually a continuous, over-complete basis (such as a coherent-state
basis\footnote{See, for instance, \cite{zurekcoherent}.} rather than a discrete one, and the very idea of a 
discretely-branching tree may be inappropriate. (I am grateful to Simon
Saunders for these observations).
\item Similarly, the process of decoherence is ongoing: branching does
not occur at discrete loci, rather it is a continual process of
divergence.
\item Even setting aside infinite-dimensional problems, the only
available method of `counting' descendants is to look at the time-evolved state
vector's overlap with the subspaces that make up the 
(decoherence-) preferred basis: when there is non-zero overlap with one of
these subspaces, I have a
descendant in the macrostate corresponding to that subspace. But the
decoherence basis is far from being precisely determined, and in
particular exactly how coarse-grained it is depends sensitively on
exactly how much interference we are prepared to tolerate between
`decohered' branches. If I decide that an overlap of $10^{-10^{10}}$ is
too much and change my basis so as to get it down to $0.9 \times 10^{-
10^{10}}$, my decision will have dramatic effects on the ``head-count'' of my descendants.
\item Just as the coarse-graining of the decoherence basis is not
precisely fixed, nor is its position in Hilbert space. Rotating it by an
angle of 10 degrees will of course completely destroy decoherence, but
rotating it by an angle of $10^{-10^{10}}$ degrees assuredly will not.
Yet the number of my descendants is a discontinuous function of that
angle; a judiciously chosen rotation may have dramatic effects on it.
\item Branching is not something confined to measurement processes. The
interaction of decoherence with classical chaos guarantees that it is
completely ubiquitous: even if I don't bother to turn on the device, I
will still undergo myriad branching while I sit in front of it. (See
\citeN[section 4]{wallacestatmech} for a more detailed discussion of this point.)
\end{itemize}

The point here is not that there is no \emph{precise} way to define the number of
descendants; the entire decoherence-based approach to the 
preferred-basis problem turns (as I argue in \citeN{wallacestructure}) upon 
the assumption that exact precision is not required. Rather, the point
is that there is \emph{not even an approximate way} to make such a
definition.

In the terminology of \citeN{wallacestructure}, the
arbitrariness of any proposed definition of number of descendants makes
such a definition neither predictive nor explanatory of any detail of the quantum
state's evolution, and so no such definition should be treated as part
of macroscopic reality. (By contrast, the macroscopic structure defined by
decoherence (which can be specified by how the weights of a family of coarse-grained projectors
in the decoherence basis change over time) \emph{is} fairly robust, and
so should be treated as real. It is only when we start probing that
structure to ridiculously high precision --- as we must do in order to
count descendants --- that it breaks down.)

So: whether or not the SU viewpoint is coherence, we are in a position
to argue that rational agents do not care about branching \emph{per se}.
From this, the erasure argument yields \textbf{equivalence}, and with it the quantum
representation theorem.

A quick way of understanding why these arguments work goes as follows.
Some physical difference between games might be:
\begin{enumerate}
\item A change to a given branch which an agent cares about;
\item A change to a given branch which an agent doesn't care about;
\item A change to the relative weights of branches; or
\item a splitting of one branch into many, all of which are
qualitatively identical for the agent's descendants in that branch.
\end{enumerate}
By \textbf{branching indifference}, (4) may be discounted. Any change of
type (1) may be incorporated into an agent's utility function without
affecting the probabilities. Changes of type (2) can be erased --- by
definition the agent doesn't care about the erasure. This only leaves
changes of type (3), which cannot be erased on pain of unitarity
violation.

\section{Conclusion}\label{conclusion}

In \citeN{decshort}, I identified four assumptions which I claimed (and
still claim!) are required for \emph{Deutsch's} proof of the probability rule
to go through: the Everett interpretation, the subjective-uncertainty
viewpoint on that interpretation, \textbf{measurement neutrality}, and a
``fairly strong set of decision-theoretic axioms''. I also argued that
\textbf{measurement neutrality} was at least plausibly a consequence of
the SU viewpoint (using roughly the argument of (the current paper's) section \ref{neutrality}.)

The present paper may be seen as an exploration of the extent to which
these assumptions can or cannot be weakened. We have found the
following:
\begin{itemize}
\item Despite the substantial reformulation of Deutsch's proof 
described above, the Everett interpretation remains crucial: as section
\ref{single} argued, the central assumption of my reformulation
(\textbf{equivalence}) is just as dependent on the Everettian assumption
as is Deutsch's own proof. Furthermore, there is no realistic prospect of 
any decision-theoretic proof which applies to (realist) interpretations
other than Everett's: all such proofs will inevitably end up requiring
us to be indifferent to which outcome is \emph{actually} going to occur,
which is absurd in any single-universe interpretation of quantum
mechanics.
\item \textbf{Measurement neutrality} \emph{per se} need not be taken as a
premise of the argument. As was argued in section \ref{direct}, it
is possible to defend \textbf{equivalence} directly, without recourse to
\textbf{measurement neutrality}.
\item The SU viewpoint is required in Deutsch's proof (I have argued)
both to defend \textbf{measurement neutrality} and to justify the
applicability of decision theory. However, the decision-theoretic axioms
which I have advanced have extremely natural justifications from the
perspective of the OD viewpoint, and sections \ref{direct}--\ref{branchingindifference} show how
someone who accepts only that viewpoint can defend \textbf{equivalence}
directly. So it seems that we are not forced to adopt the SU
viewpoint, at least not in order to prove the probability rule.
\item Deutsch's decision-theoretic axioms are far stronger than is
strictly necessary. The notion of a minimally rational preference
ordering discussed in section \ref{axioms} is much weaker, yet still
sufficiently strong to derive the quantum representation theorem. Hence,
criticisms of Deutsch's program based on the specifics of his decision
theory appear to be beside the point.
\end{itemize}

Given the enormous contribution that decoherence theory has made to the problem
of defining a preferred basis, quantitative probability is arguably the last major 
obstacle confronting the Everett interpretation. In my view the evidence is now quite
strong that the decision-theoretic
strategy which Deutsch suggested is able to solve the problem; if so,
the significance for Everett's program is hard to overstate.

\subsection*{Acknowledgements}

For useful conversations and detailed feedback at various points over
the evolution of this paper, I would like to thank Harvey Brown, Adam Elga, Hilary Greaves, Peter
Lewis, David Papineau, and especially Jeremy Butterfield and Simon
Saunders.

\end{document}